\documentclass[a4paper]{article}
\usepackage{graphicx}
\usepackage{hyperref}
\usepackage{a4wide}
\usepackage[utf8]{inputenc}

\newtheorem{observation}{Observation}

\begin{document}

\begin{center}
{\Large \bf
Do you trade with your friends\\
\smallskip
or become friends with your trading partners?\\
\smallskip
A case study in the $\tilde{G}1$ cryptocurrency
}\\
\medskip
Nicolas Gensollen
and
Matthieu Latapy
\end{center}

\begin{abstract}
We study the interplay between social ties and financial transactions made through a recent cryptocurrency called $\tilde{G}1$. It has the particularity of combining the usual transaction record with a reliable network of identified users. This gives the opportunity to observe exactly who sent money to whom over a social network. This social network is a key piece of this cryptocurrency, which therefore puts much effort in ensuring that nodes correspond to unique, well identified, real living human users, linked together only if they met at least once in real world. Using this data, we study how social ties impact the structure of transactions and conversely. We show that users make transactions almost exclusively with people they are connected with in the social network. However, they tend to build social connections with people they will never make transactions with.
\end{abstract}

%%%%%%%%%%%%%%%%%%%%%%%%%%%%%%%%%%
%      SECTION I - INTRODUCTION
%%%%%%%%%%%%%%%%%%%%%%%%%%%%%%%%%%
%
\section{Introduction}
\label{sec:introduction}

Since the introduction of the blockchain in 2008 \cite{Satoshi}, the number and diversity of applications relying on this technology has been continuously growing \cite{Al-Saqaf} \cite{Hileman}. Although Bitcoin is still a benchmark cryptocurrency \cite{Hileman} \cite{Gohwong}, many new currencies relying on different kinds of blockchains have been introduced since then \cite{Gohwong}.

Understanding the growth and popularity of cryptocurrencies is non-trivial, but social ties seem to play a key role. This is not surprising since currencies enable transactions between individuals, and for these transactions to actually happen, some kind of pre-existing tie is often required. In the early age of cryptocurrencies, actually, specific platforms were forming the bulk of these media. Cryptocurrencies then relied on social media to boost their growth and facilitate interactions.

%It is not trivial to understand what contributes to the growth and popularity of cryptocurrencies and other blockchain-based applications, but social ties seem to play a key role. This is hardly a surprise since a currency's main purpose is to enable transactions between individuals, and for these transactions to actually happen, some kind of pre-existing tie is often required. In the early age of cryptocurrencies, specific forums and exchange platforms were forming the bulk of these media. Blockchain based applications were thus relying on social media to boost their growth and facilitate interactions. Nowadays, technologies using a blockchain to provide decentralized integrity are often considered to be the next step in the evolution of social media.

%The Steem blockchain for example was developed and realeased in the end of March 2016 \cite{Steem}. Shortly after, the decentralized and censorship-resistant social platform Steemit was launched as the first application running on the Steem protocol \cite{Steemit}. The key idea behind Steemit is that members of the community get rewarded for sharing content, the retribution increasing with the content's popularity. A vote for a given content may thus be seen as some kind of transaction occurring between two members of a social network through a particular publication.

%While platforms like Steemit use both blockchains and cryptocurrencies to change the way people interact on social media, other platforms use social interactions as a way to strengthen the robustness of the system.
The $\tilde{G}1$ \cite{g1} cryptocurrency goes further: it relies on social ties to strengthen the robustness of the system. It maintains an accurate network of identified users with reliable social ties, and use it for monetary growth. This offers a unique opportunity to study the interplay between financial transactions and social ties between human beings.

% where the monetary growth depends on a network of identified users, is such a system. As we will explain in more details in section \ref{sec:g1}, $\tilde{G}1$ offers a fascinating dataset to study the interplay between financial transactions and social ties between identified human beings. 

%%%%%%%%%%%%%%%%%%%%%%%%%%%%%%%%%%
%      I.1 - RELATED WORK
%%%%%%%%%%%%%%%%%%%%%%%%%%%%%%%%%%
%
\subsection{Related Work}
\label{subsec:related_work}

The literature on social interactions on the one hand, and financial transactions on the other hand is extremely vast and spans several domains such as economy, sociology, network science, psycology, or finance to name only a few. Yet, scientific works studying how these phenomena co-evolve remain scarce. A key reason is the lack of appropriate data sources: financial transactions are considered as sensitive data, and rarely made public; even when they are, interactions are anonymized. In \cite{Martens} for instance, the authors use real but anonymized transaction records to infer a pseudo-social network of users in which two users are connected if they transfered money to the same entity. Then, they use this pseudo-social network for social targeting and show that it performs better at identifying buyers than targeting based on traditional models.

Pioneer work studying both social ties and the way people make transactions can be found in sociology. In \cite{Zelizer} for example, the authors propose to split payements in three categories: gifts, entitlements, and compensations, and show that each category corresponds to a specific set of social relationships and systems of meanings. The recent development of online social networks and the increasing availability of data gave a new boost to the study of social ties. Among this vast literature, a subset of contributions studies how social ties can be predicted from other data sources. For example \cite{Crandall} uses the proximity, in space and time, of geo-tagged photographs over the Flickr social network to infer the likelihood of a social tie between users. This paper shows that this probability increases by orders of magnitudes as the number of co-locations increases. In \cite{Hristova}, the authors explore the combined effect of multiple social networks for link prediction. More precisely, they represent social interactions as a multiplex network where each layer represents a specific social platform, and they show how this additional information can be used to improve link prediction. In \cite{Khosravi}, the authors investigate link strength prediction in a social network based on social transactions (likes, comments, etc). They propose a new type of multiple-matrix factorization model for incorporating a transaction matrix between users, and test their method on \textit{Cloob} \cite{Cloob}, a popular Iranian social network where users can rate their friendship relationships.

Related works basically either start from a transaction record to infer underlying social ties, or start from a social network and try to predict new or missing links based on various features. To the best of our knowledge, there is no previous work studying financial transactions and social interactions simultaneously from a reliable data source. The recent development of cryptocurrencies is creating new opportunities for this kind of studies. Contrary to transactions relying on usual payment methods, blockchain based transactions are public and can be analyzed freely as long as the blockchain itself is public. In \cite{Kondor}, for instance, the authors extracted the transactions from the Bitcoin blockchain and reconstructed the network of transactions. They provide a graph-based analysis of this network and show that linear preferential attachment drives its growth. In \cite{Popuri} the authors also study the network of transactions of both Bitcoin and Litecoin, while the authors of \cite{Maesa} recently studied the structure of the Bitcoin users graph, exhibiting a bow tie like structure between its components. In \cite{Kim}, the authors analyzed user comments in online communities of Bitcoin, Ethereum, and Ripple to predict the price and number of transactions in these cryptocurrencies.

The main limitation is often that, in most of these systems, the public keys used as wallets for transactions are used only once, i.e. users are encouraged to create a set of cryptographic keys for each transaction as a way to preserve anonymity. A direct consequence for our purpose is that there is no obvious way to link real users to the set of keys they used to make transactions \cite{Meiklejohn} \cite{Cazabet}. Some heuristics have been proposed to tackle this challenge \cite{Meiklejohn} \cite{Cazabet} but they mostly work for large users and it is difficult to assess their reliability. In addition, even though users were identified properly, the underlying social ties would still be unknown.

%%%%%%%%%%%%%%%%%%%%%%%%%%%%%%%%%%
%      I.2 - OUR CONTRIBUTION
%%%%%%%%%%%%%%%%%%%%%%%%%%%%%%%%%%
%
\subsection{Our contribution}
\label{subsec:contribution}

In this paper, we study a specific cryptocurrency which offers both a ledger of transactions and of social bounds between identified human beings. This means that we can tell exactly who sent money to whom and when. Our main objective is to understand the interplay between these transactions and the social ties. More precisely, we first wish to understand whether users start making transactions before creating a tie, or if they tend to make transactions with people they are already friends with. Going further, we wish to study the different neighborhood structures and their evolution over time. We will tackle questions such as: Are my transaction partners the same as my friends? How do my friends exchange between them compared to my transaction partners? Are my friends and transaction partners more and more homogeneous over time?

As we will see in section \ref{sec:methods}, although the data is rather simple at first glance, the proper modeling of interactions is still a challenge and no unique, commonly accepted approach exists. We leverage here the recently introduced stream graph model, which captures both the time and structural components of data \cite{stream_graphs} \cite{weighted_stream_graphs}. We will start our analysis with basic metrics targetting the questions above and we will introduce new stream graph concepts as we need them in the analysis. 

This paper is organized as follows. Section \ref{sec:g1} introduces the $\tilde{G}1$ cryptocurrency and explains the main ideas and mechanisms behind it. In section \ref{sec:methods}, we present the dataset under study and show how the stream graphs model interactions from this dataset. In section \ref{sec:overview}, we use time series and static graph concepts to gain a first insight on the global structure and dynamics of the system. In section \ref{sec:link_based} we consider both time and structure together but stay at a basic link level in order to understand the interplay between social ties and transactions. Finally, in section \ref{sec:neighborhoods} we use more complex stream graph concepts mixing time and structure in order to investigate these questions further.

%%%%%%%%%%%%%%%%%%%%%%%%%%%%%%%%%%%%%%%%%%%
%      SECTION II - THE G1 CRYPTOCURRENCY
%%%%%%%%%%%%%%%%%%%%%%%%%%%%%%%%%%%%%%%%%%%
%
\section{The $\tilde{G}1$ cryptocurrency}
\label{sec:g1}

%%%%%%%%%%%%%%%%%%%%%%%%%%%%%%%%%%
%      II.1 - A LIBRE CURRENCY
%%%%%%%%%%%%%%%%%%%%%%%%%%%%%%%%%%
%
\subsection{A \textit{libre} currency}
\label{subsec:libre_currency}

$\tilde{G}1$ \cite{g1} is a cryptocurrency introduced in France in March 2017. Contrary to most other cryptocurrencies, the growth of the $\tilde{G}1$ monetary mass is not related to mining but to the number of users, and it is distributed evenly between them. Each member receives a share of the monetary growth every day as the \textit{universal dividend} (UD). This share should not be considered as a salary or as a reward: there is no work required from the members to earn their UD (although they have some responsibilities as we will see). The universal dividend simply comes from the fact that all individuals are considered equal in terms of currency creation.  $\tilde{G}1$ is indeed considered to be the first "\textit{libre}" cryptocurrency (which means \textit{free} in French), and which properties were developed in the relative theory of money \cite{TRM}.

\textit{Libre} currencies have two key properties which differentiate them from other currencies: a spatial and a temporal symmetry. The spacial symmetry means that all individuals of a given generation co-create the same amount of new currency units, while the temporal symmetry prevents a generation of users to be favored over another by the currency creation process. In $\tilde{G}1$, the monetary growth rate, and therefore the UD's value, evolve according to the population's size and are updated every six months \cite{g1}. 

It is easy to see that most currencies, independantly of their use of the blockchain, are not \textit{libre}: some individuals benefit from the monetary creation at the expense of others. In Bitcoin for example, early adopters managed to mine most of the finite monetary mass leaving only crumbs to the other users \cite{Kondor} \cite{Rose}. With fiat currencies, creation of new units is often a priviledge given to states and remains obscur to most citizens. One of the main objectives of $\tilde{G}1$ and the relative theory of money \cite{TRM} is to show that a currency can fulfill its purpose of enabling transactions of goods, while preserving fairness and equity in terms of monetary creation.

%%%%%%%%%%%%%%%%%%%%%%%%%%%%%%%%%%%%%%%%%
%      II.2 - IDENTIFICATION OF MEMBERS
%%%%%%%%%%%%%%%%%%%%%%%%%%%%%%%%%%%%%%%%%
%
\subsection{Identification of members}
\label{subsec:members}

The whole $\tilde{G}1$ system relies on its members and their identities. Indeed, each account must be attached to a well identified living human being with no other account in the system. Otherwise anybody would be able to create multiple accounts and earn multiple UDs every day, which would undoubtly trigger the collapse of the whole system. We will see later that, although they cannot own a member account, non-human entities like corporations or associations can still participate to the transaction network.

In order to identify its members, $\tilde{G}1$ maintains a web of trust (WoT) between them, in addition to the usual transaction record. A link from person $a$ to person $b$ in the WoT is called a \textit{certification} and means that $a$ certifies that $b$ is a real living human being, with no other account in the system. These certifications are written in the blockchain and have a finite validity (currently set to two years), meaning that members have to renew their certifications to maintain their status. $\tilde{G}1$ implements a set of rules to ensure robustness of the WoT against attacks. For example, a new member has to obtain at least five certifications from already existing members to join the system and contribute to the currency creation process. Moreover, any member has to be within a small distance to a  member with enough certifications given and received for its identity to be considered as safe. As we will see, these rules shape the WoT in a rather clustered network with a small diameter. 

Although robutness of the WoT against sybil attacks is an interesting topic on its own, we will not consider it here. In this paper, we assume that the WoT is composed of real unique human beings and that certification links accurately reflect social ties between them. Note that each new member agrees and signs a charter explaining the responsibilities of a member. Among them, a member has to be sure of the validity of a new account before giving a certification. This means that a new user wishing to become a member has to meet other members in real life at least once. Although this can be a difficult thing to do, the community organizes social events to welcome new users. A certification is thus not necessarily a strong social tie per se, but it is supposed to exist only between human beings who have met at least once in real life, which might be more than one can expect in other online social networks.

%%%%%%%%%%%%%%%%%%%%%%%%%%%%%%%%%%%%%%
%      II.3 - FINANCIAL TRANSACTIONS
%%%%%%%%%%%%%%%%%%%%%%%%%%%%%%%%%%%%%%
%
\subsection{Financial transactions}
\label{subsec:transactions}

In addition to the WoT, $\tilde{G}1$ enables financial transactions which are also written in the blockchain like for usual cryptocurrencies. There is no obligation to be a member to make transactions. $\tilde{G}1$ has indeed a clear distinction between a member account, which participates in the monetary growth and receives one UD per day (see section \ref{subsec:members}), and a wallet, which is nothing more than a public key that can send and receive money. Anybody can create one or several wallets, and transactions can be made freely between wallets and members.

We will come back to this in section \ref{subsec:objects}, but this distinction enables us to split the stream of transactions in different substreams according to source and destination types. One of these substreams is composed only of transactions between identified members, such that we know exactly who sent what to whom and when within this stream. This is already much more than one can hope for in other cryptocurrencies like Bitcoin where transactions occur between one time only keys. But, we also have, thanks to the WoT, the social relationships between the entities making these transactions. This makes $\tilde{G}1$ a unique dataset to study the interplay between social ties and financial interactions.

%%%%%%%%%%%%%%%%%%%%%%%%%%%%%%%%%%%%%%%%%%%%%%%%%%%%%%%%%
%      SECTION III - DATASET AND STREAM GRAPH MODELING
%%%%%%%%%%%%%%%%%%%%%%%%%%%%%%%%%%%%%%%%%%%%%%%%%%%%%%%%%
%
\section{Dataset and stream graph modeling}
\label{sec:methods}

All the data we consider here is extracted from the $\tilde{G}1$ blockchain which is publicly available \cite{g1_blockchain}. For our purpose, the blockchain contains three key information: the \textit{identities}, the \textit{certifications}, and the \textit{transactions}. Identities associate a public key (which we also call a \textit{wallet}) to a user name and represent therefore the members of the system. As explained in section \ref{subsec:members}, a certification is a directed interaction between a member and a public key, while a transaction is a directed weighted interaction between two public keys. From the blockchain, we extracted a stream of certifications, which is basically a sequence of $(t,u,v)$ triplets, meaning that entity $u$ certifies entity $v$ at time $t$, and a stream of transactions, which is a sequence of $(t,u,v,a)$ quadruplets, meaning that entity $u$ sends an amount $a$ to entity $v$ at time $t$. 

Although simple in appearance, this data has both a structural and temporal component which makes its analysis far from trivial. Indeed, relying on classical methods such as static graphs or time series means that the data is somehow simplified and that information is lost in the process. In this paper, we propose to use \textit{stream graphs}, rencently introduced in \cite{stream_graphs} and \cite{weighted_stream_graphs}, to study these streams, and the dependencies between them. We provide first some basic notions and notations related to the analysis of interaction data using the stream graph formalism, and we encourage the interested reader to have a look at references \cite{stream_graphs} and \cite{weighted_stream_graphs} for a more detailed introduction to stream graphs.

% FIGURE 1
%
\begin{figure}
\begin{center}
\includegraphics[scale=.3]{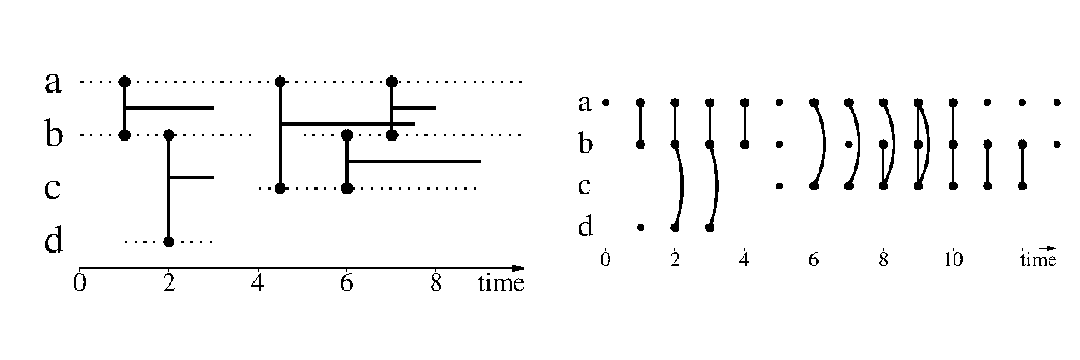}
\caption{\textbf{Left} - Example of a stream graph with $ T = \left[0,10 \right] $, $V = \left\{ a,b,c,d \right\}$, $W = \left[0,10 \right] \times \left\{ a \right\} \cup \left( \left[ 0,4 \right] \cup \left[ 5,10 \right] \right) \times \left\{ b \right\} \cup \left[ 4,9 \right] \times \left\{ c \right\} \cup \left[ 1,3 \right] \times \left\{ d \right\} $, and $ E = \left( \left[ 1,3 \right] \cup \left[ 7,8 \right] \right) \times \left\{ ab \right\} \cup \left[ 4.5,7.5 \right] \times \left\{ ac \right\} \cup \left[ 6,9 \right] \times \left\{ bc \right\} \cup \left[ 2,3 \right] \times \left\{ bd \right\} $. In other words, $ T_a = \left[0,10 \right] $, $ T_b = \left[0,4\right] \cup \left[5,10 \right] $, $T_c = \left[ 4,9 \right] $, $T_d = \left[1,3\right]$, $ T_{ab} = \left[ 1,3 \right] \cup \left[7,8 \right] $, $ T_{ac} = \left[ 4.5, 7.5 \right] $, $ T_{bc} = \left[6,9 \right] $, $ T_{bd} = \left[ 2,3 \right] $, and $ T_{ad} = T_{cd} = \emptyset $.  \textbf{Right} - An example of stream graph in discrete time. It is defined by $T=\left[0,13 \right] \subseteq N $, $V=\left\{a,b,c,d\right\}$, $T_a = T$, $T_b = \left\{1, 2, 3, 4, 5, 7, 8, 9, 10, 11, 12, 13 \right\} $, $T_c = \left[5,12\right]$, $T_d = \left[1,3\right]$, $T_{ab} = \left[1,4\right] \cup \left[9,10 \right]$, $T_{ac} = \left[6,9\right]$, $T_{bc} =\left[8,12\right]$, $T_{bd} = \left[2,3 \right]$, and $T_{ad} = T_{cd} = \emptyset $.}
\label{fig:streamgraph}
\end{center}
\end{figure}

%%%%%%%%%%%%%%%%%%%%%%%%%%%%%%%%%%
%      III.1 - STREAM GRAPHS
%%%%%%%%%%%%%%%%%%%%%%%%%%%%%%%%%%
%
\subsection{Stream graphs}
\label{subsec:streamgraphs}

A stream graph $S$ is defined as a quadruplet $(T,V,W,E) $ where $T$ is the set of time instants, $V$ is a finite set of nodes, $W \subseteq T \times V$ is a set of temporal nodes, and $E \subseteq T \times V \otimes V$ is a set of links such that $\left( t,uv \right) \in E $ implies $ \left( t,u \right) \in W $ and $ \left( t,v \right) \in W $. The left plot of figure \ref{fig:streamgraph} shows an example of a simple stream graph with four nodes $a$, $b$, $c$, and $d$ (node labels are represented on the y-axis) over a time period going from $t=0$ to $t=10$ (time is represented on the x-axis). The nodes' presence is represented as horizontal dotted lines. For example, on the left plot of figure \ref{fig:streamgraph}, node $b$ is present between $t=0$ and $t=4$, and then between $t = 5$ and $t=10$. An interaction between two nodes at a given time is represented as a vertical bold line connecting them. Depending on the dynamics of the system, interactions can have a duration or can be viewed as discrete. For example, an SMS can be seen as a discrete interaction between two phone numbers, while a phone call would most likely be modeled as a link with duration. When a link has a duration, it is represented as an horizontal bold line. On the left plot of figure \ref{fig:streamgraph} for example, nodes $a$ and $b$ interact from $t=1$ to $t=3$, and from $t=7$ to $t=8$. Note also that the set of time instants $ T $ can be continuous or discrete. The right plot of figure \ref{fig:streamgraph} shows a simple example of a stream graph in discrete time. A few stream graph concepts differ slightly from continuous to discrete time (mostly integrals becoming sums and sets becoming intervals) but most concepts apply directly to both cases.

%%%%%%%%%%%%%%%%%%%%%%%%%%%%%%%%%%
%      III.2 - MODELING
%%%%%%%%%%%%%%%%%%%%%%%%%%%%%%%%%%
%
\subsection{Modeling certifications and transactions using stream graphs}
\label{subsec:modeling}

% Questions ML: directed? def of $W$??

We denote by $ V $ the set of all public keys interacting at least once in the certification or transaction streams. A subset $ M \subseteq V $ of these public keys belongs to members of the system and is available through the identities written in the blockchain. The set $ A = V \setminus M $ contains all public keys which were involved in at least one transaction and do not belong to a member, which we will call \textit{anonymous wallets} in the following.

We denote by $ \mathcal{C} = \left( T_{\mathcal{C}}, M, W_{\mathcal{C}}, E_{\mathcal{C}} \right) $ the stream of certifications between members: $ T_{\mathcal{C}} = \left[ 1488987127, 1555054577 \right] $ is the time interval going from the first certification given (on 2017-03-08 15:32:07) to the latest certification considered in this study (on 2019-04-12 07:36:17); $ W_{\mathcal{C}} $ is the set of temporal nodes involved in $ \mathcal{C} $; and $  E_{\mathcal{C}} $ is the set of certification links.

Similarly, we denote by $ \mathcal{T} = \left( T_{\mathcal{T}}, V, W_{\mathcal{T}}, E_{\mathcal{T}} \right) $ the stream of transactions: $ T_{\mathcal{T}} = \left[ 1488990898, 1555052722 \right] $ is the time interval going from the first transaction (on 2017-03-08 16:34:58) to the latest transaction considered in this study (on 2019-04-12 07:05:22); $ W_{\mathcal{T}} $ is the set of temporal nodes involved in $ \mathcal{T} $; and $  E_{\mathcal{T}} $ is the set of transaction links. 

When the currency was launched in $2017$, the initial block was composed of $49$ members who certified each other to initialize the WoT with $551$ certifications. New users then joined the system by becoming members and/or by making transactions in $\tilde{G}1$. This means that nodes are not always present in the streams, they appear at some point in time, that we call their \textit{birth}, and they may exit the system at a later time. Over the period of study (2017-2019), $\mathcal{C}$ contains $2128$ nodes (i.e $ \left| M \right| = 2128 $) and $18765$ distinct certifications, while $\mathcal{T}$ contains $38296$ transactions between $3872$ public keys.

Recall that a certification between two members is valid for two years, such that a link in $\mathcal{C}$ can be seen as a link with a duration of two years. Since $\tilde{G}1$ is still a young cryptocurrency, we currently have only two years of data available (2017-2019), such that the very slow link dynamics in $\mathcal{C}$ is not very interesting yet: links basically appear at some point and stay active until the end of the stream. Transactions on the other hand form a stream of links without a clearly defined duration. In this paper, we will use a discrete time representation for $\mathcal{T} $ in order to avoid technical difficulties which can occur when dealing with discrete interactions in continuous time. Note however, that we could have chosen an arbitrary link duration (the time interval between two blocks in the blockchain for example) for a transaction in order to stay in continuous time.

%%%%%%%%%%%%%%%%%%%%%%%%%%%%%%%%%%%%%%%%%%%%%
%      III.3 - USEFUL OBJECTS AND CONCEPTS
%%%%%%%%%%%%%%%%%%%%%%%%%%%%%%%%%%%%%%%%%%%%%
%
\subsection{Useful objects and concepts}
\label{subsec:objects}

As explained before, a public key can be linked to a member account or to an anonymous wallet such that we can define two subsets of $W_{\mathcal{T}}$ (called \textit{clusters}), one containing only members involved in transactions, and the other containing only anonymous wallets involved in transactions. We can then define the set of links between nodes involved in a cluster as the \textit{substream} induced by this cluster (see \cite{stream_graphs} for more details). Transactions can thus occur between identified members, between anonymous wallets, or between members and anonymous wallets such that we can split $ \mathcal{T} $ in four non-overlapping substreams: $ \mathcal{T}_{MM} $ (transactions between identified members), $ \mathcal{T}_{MA} $ (transactions from a member to an anonymous wallet), $\mathcal{T}_{AM} $ (transactions from an anonymous wallet to a member), and $\mathcal{T}_{AA} $ (transactions between anonymous wallets).

Since stream graphs encode both time and structure, it is natural to define the \textit{graph induced} by a stream and the \textit{activity} of a stream. The graph induced by a stream $S = \left( T,V,W,E \right) $ is a static graph $G \left( S \right) = \left( V, \bar{E} \right) $ in which two nodes are linked if they interacted at least once in the stream, i.e. $ \bar{E} = \left\{ \left(u,v \right),\ \exists \left( t,uv \right) \in E \right\} $. Note that the WoT is then nothing more than the graph induced by $\mathcal{C}$. The activity of a stream is often defined as the time serie of its link activity, i.e. the number of active links as a function of time.

%%%%%%%%%%%%%%%%%%%%%%%%%%%%%%%%%%%%%%%%%%
%      SECTION IV - OVERVIEW OF C and T
%%%%%%%%%%%%%%%%%%%%%%%%%%%%%%%%%%%%%%%%%%
%
\section{Overview of certification and transaction streams}
\label{sec:overview}

\begin{figure}
\centering
\includegraphics[scale=.4]{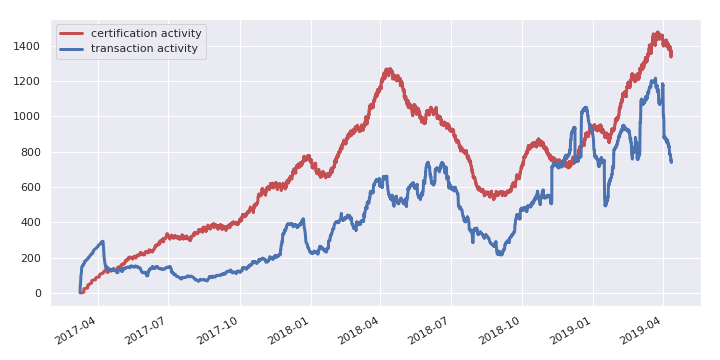}
\caption{30 days rolling sum of the activities of the certification stream $ \mathcal{C} $ (red curve) and the transaction stream between members $ \mathcal{T}_{MM}$ (blue curve).}
\label{fig:activity}
\end{figure}

In this section, we study general properties of $ \mathcal{C} $ and $ \mathcal{T} $ in order to gain a better understanding of their structures and dynamics. We start by studying the activities of the certification stream $ \mathcal{C}$, and of the transaction stream between members $ \mathcal{T}_{MM} $. Figure \ref{fig:activity} shows the 30 days rolling sum of both activities, the red curve showing the certification activity while the blue curve shows the transaction activity. As can easily be seen, both activities are strongly correlated and follow very similar trends. We can observe a first growth period going from March 2017 to April 2018 before a strong decrease in activity until early October 2018. Since then, both transaction's and certification's activities are increasing with a little more volatility on the transactions' side.

% FIGURE 2
%

% OBSERVATION 1
%
\begin{observation}
The certification dynamics is strongly correlated the one of transactions among members.
\label{obs:1}
\end{observation}

\begin{figure}
\centering
\includegraphics[scale=.3]{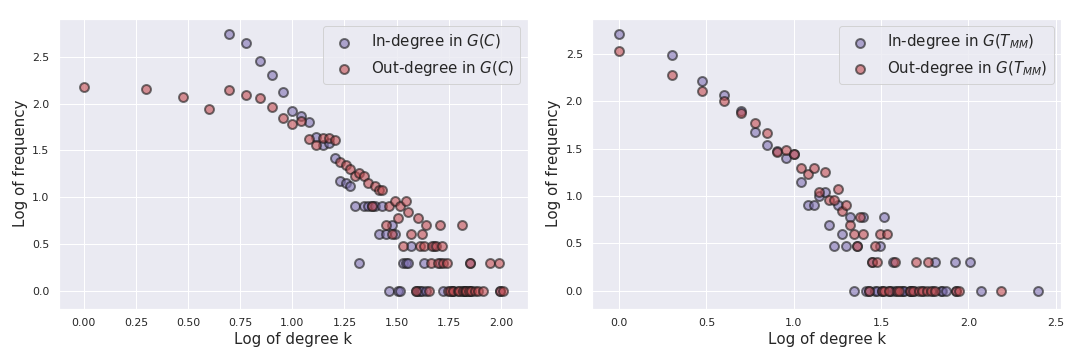}
\caption{\textbf{Left -} In-degree (in blue) and out-degree (in red) distributions of $G \left( \mathcal{C} \right) $. \textbf{Right -} In-degree (in blue) and out-degree (in red) distributions of $ G \left( \mathcal{T}_{MM} \right) $. }
\label{fig:degree_distribution}
\end{figure}

If the activity of a stream gives a rough idea of the global dynamics, the induced graph can be used to visualize a simplification of the stream's structure. By projecting the interactions on a static graph, we obtain two directed graphs $ G\left( \mathcal{C} \right) $ and $G \left( \mathcal{T}_{MM} \right) $. In figure \ref{fig:degree_distribution}, we show the in and out degree distributions of each graph. The left plot shows the distributions for $ G\left( \mathcal{C} \right) $ while the right plot shows the distributions for $G \left( \mathcal{T}_{MM} \right) $. Both graphs clearly display a heavy tail degree distribution meaning that some members have much larger certification and transaction neighborhoods than the majority of members.

% OBSERVATION 2
%
\begin{observation}
Both the cortification graph and the member transaction graph have a heterogeneous degree distribution.
\label{obs:2}
\end{observation}

\begin{figure}
\begin{center}
\includegraphics[scale=.3]{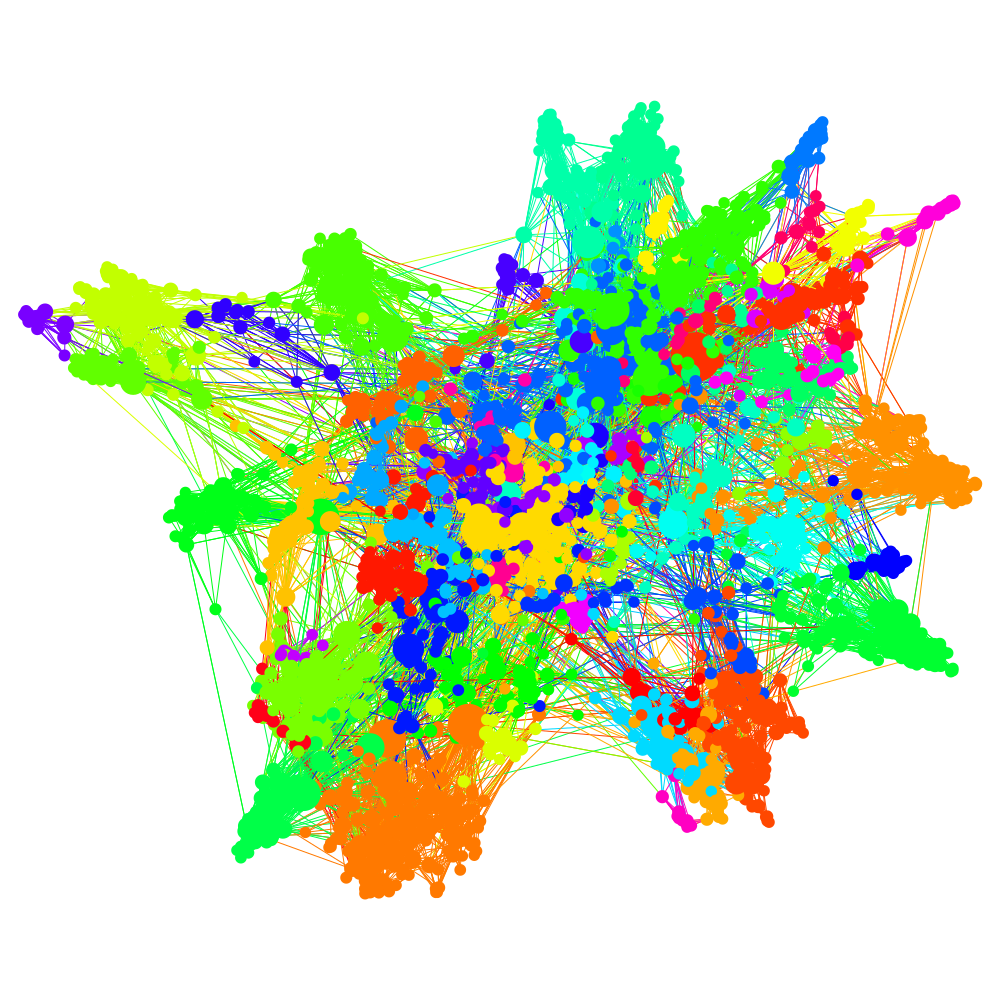}
\end{center}
\caption{Visualization of the WoT obtained from the $\tilde{G}1$ visualization tool \cite{WoT_visual}. Nodes represent members while links represent certification relationships. Colors show communities.}
\label{fig:wot}
\end{figure}

Figure \ref{fig:wot} shows a representation of $G \left( \mathcal{C} \right) $ and was obtained from \cite{WoT_visual}. Note that, although this is not the focus of this paper, $G \left( \mathcal{C} \right) $ seems to have an interesting community structure.

% FIGURE 3
%

\begin{figure}
\begin{center}
\includegraphics[scale=.3]{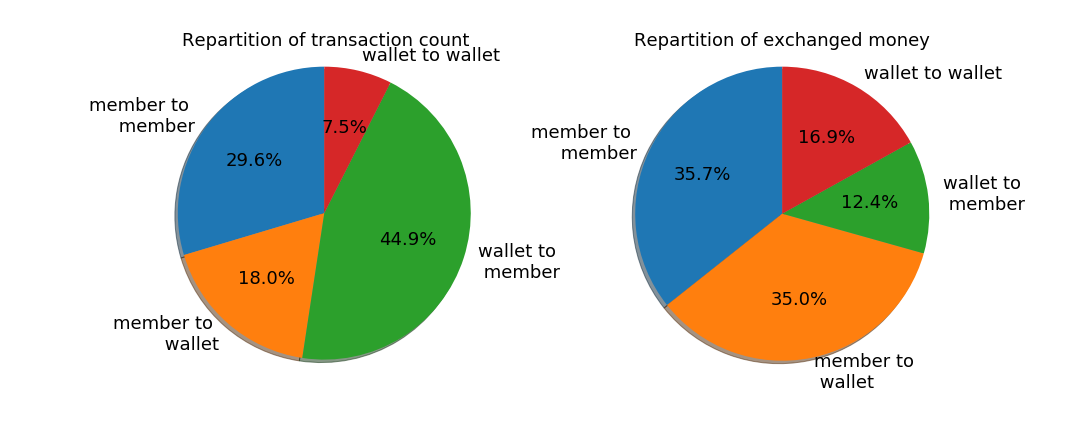}
\caption{The transaction stream $ \mathcal{T} $ can be divided in four substreams: transactions between identified members $\mathcal{T}_{MM} $, between a member and an anonymous wallet $ \mathcal{T}_{MA} $, between an anonymous wallet and a member $ \mathcal{T}_{AM} $, and between anonymous wallets $ \mathcal{T}_{AA} $. \textbf{Left -} Repartition of the transactions between the different transaction substreams. \textbf{Right -} Repartition of the exchanged money between the different transaction substreams. Note that, although $ \mathcal{T}_{AM} $ represents about $45\%$ of $ \mathcal{T}$ in terms of number of transactions, it only adds up to $12.4\%$ of the exchanged volume.}
\label{fig:repartition}
\end{center}
\end{figure}

We focus now on the stream of transactions $ \mathcal{T} $. Recall that we defined in section \ref{subsec:objects} four distinct substreams depending on the origin/destination types of the transactions: $\mathcal{T}_{MM} $, $\mathcal{T}_{MA} $, $\mathcal{T}_{AM} $, and $\mathcal{T}_{AA} $. Figure \ref{fig:repartition} shows the repartition of the transactions between these four substreams as well as the repartition in terms of the total exchanged volume. As can be seen, transactions between identitfied members (i.e. belonging to $\mathcal{T}_{MM} $) represent only $30\%$ of $\mathcal{T}$ while transactions between an anonymous wallet and a member (i.e. belonging to $\mathcal{T}_{AM} $) represent almost $45\%$ of $ \mathcal{T}$ and only $12.4\%$ in terms of exchanged money, meaning that there are a lot of small transactions going from anonymous wallets to member accounts. One possible reason for this asymmetry is linked to the way miners are retributed in $\tilde{G}1$. Recall that money creation is done by the members themselves and that it has nothing to do with mining, such that miners are doing the work for free by design. In order to compensate the miners for this voluntary work, there is a special wallet in $\tilde{G}1$, called \textit{Remuniter}, which receives donations and uses them to retribute the miners for the work they have done. This very specific wallet is therefore a very large hub in $\mathcal{T}$ and stands for almost $40\%$ of all transactions.

% OBSERVATION 3
%
\begin{observation}
A large fraction of all transactions is related to miner retribution.
\label{obs:3}
\end{observation}

Note that, thanks to this specific wallet, we can easily identify the miners of $\tilde{G}1$ as the  outgoing neighbors of \textit{Remuniter} in $\mathcal{T}_{AM}$ (i.e. all members who received money from \textit{Remuniter}). At the time we downloaded the blockchain, there was $158$ miners among the identified members of the system.

% FIGURE 4
%

% FIGURE 5
%

\begin{figure}
\centering
\includegraphics[scale=.32]{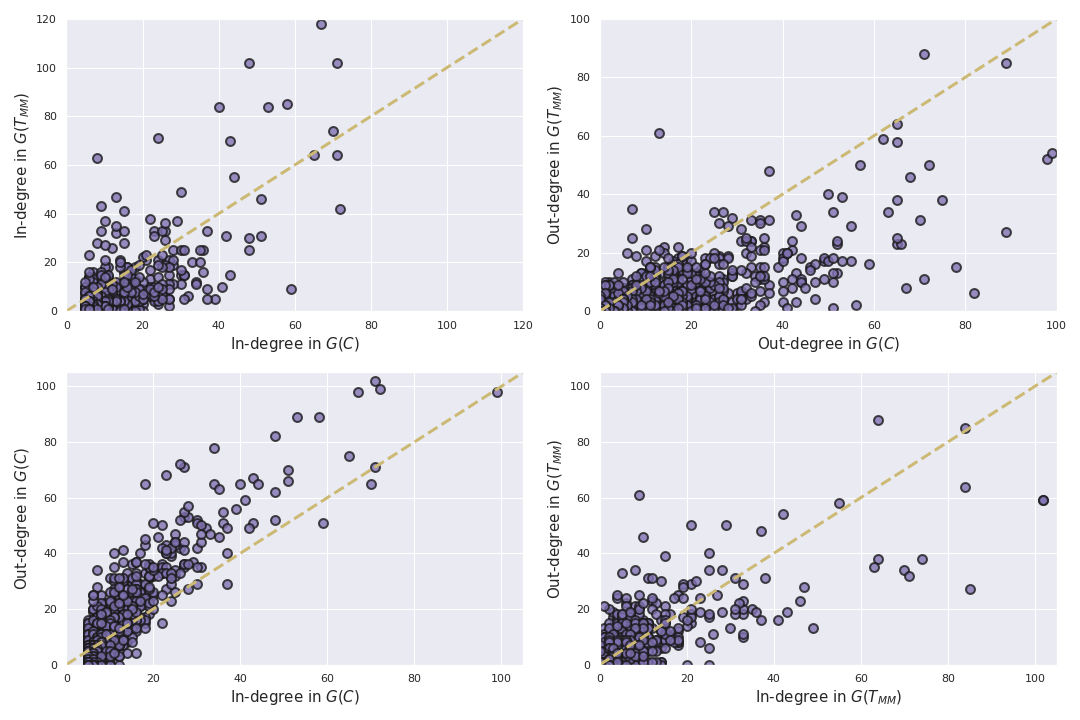}
\caption{\textbf{Top Left -} In-degree in $G \left( \mathcal{T}_{MM} \right) $ vs. in-degree in $ G \left( \mathcal{C} \right) $. \textbf{Top Right -} Out-degree in $G \left( \mathcal{T}_{MM} \right) $ vs. out-degree in $ G \left( \mathcal{C} \right) $. \textbf{Bottom Left -} Out-degree in $G \left( \mathcal{C} \right) $ vs. in-degree in $ G \left( \mathcal{C} \right) $. \textbf{Bottom Right -} Out-degree in $G \left( \mathcal{T}_{MM} \right) $ vs. in-degree in $ G \left( \mathcal{T}_{MM} \right) $.}
\label{fig:homophily}
\end{figure}

Figure \ref{fig:homophily} shows the correlations between the nodes' degrees in $ G \left( \mathcal{C} \right) $ and $ G \left( \mathcal{T}_{MM} \right) $. First, note that there is no node with an in-degree smaller than $5$ in $ G \left( \mathcal{C} \right) $ because of the minimum number of certifications required to become a member. The top right subplot shows that members tend to give more certifications than they tend to initiate transactions, especially for high degree values. The bottom left subplot shows that nodes in-degree and out-degree values in $ G \left( \mathcal{C} \right) $ are strongly correlated, meaning that people who receives a lot of certifications tend to also give a lot of them, and tend to actually give more certifications than they received.

% OBSERVATION 4
%
\begin{observation}
Members tend to give more certifications than they initiate new transactions. In addition, there is a strong correlation in the number of certifications given and received per member.
\label{obs:4}
\end{observation}

% FIGURE 6
%

%%%%%%%%%%%%%%%%%%%%%%%%%%%%%%%%%%%%%%%%%%
%      SECTION V - NEW LINKS IN C and T
%%%%%%%%%%%%%%%%%%%%%%%%%%%%%%%%%%%%%%%%%%
%
\section{How do new certifications and transactions appear between members?}
\label{sec:link_based}

Our main focus in this paper is to study how certifications impact transactions and vice-versa such that we will focus on the certification stream $ \mathcal{C} $ and the transaction stream restricted to the identified members $ \mathcal{T}_{MM} $. In this section we use a simple link-based approach to understand how new links appear in these two streams. More specifically, we wish to understand if a relationship between two members tends to exist in both streams and if it rather starts with a certification (i.e. with a social tie) or through transactions.

\begin{figure}
\begin{center}
\includegraphics[scale=.37]{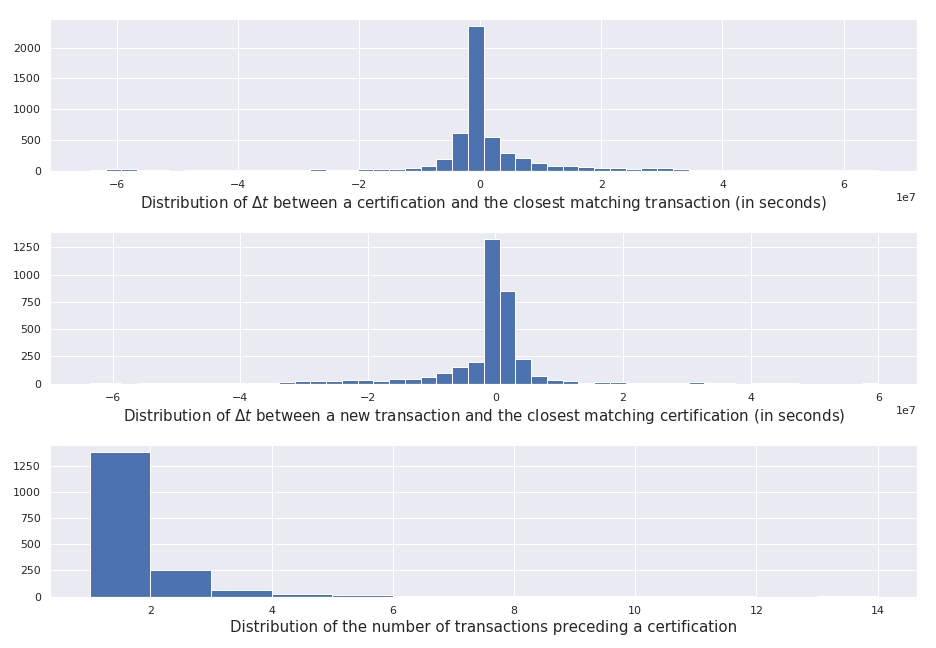}
\caption{\textbf{Top} - Distribution of time interval $\Delta t$ (in seconds) between a certification in $\mathcal{C}$ and the closest matching transaction in $ \mathcal{T}_{MM} $. \textbf{Middle -} Distribution of the time interval $ \Delta t $ (in seconds) between a new transaction in $ \mathcal{T}_{MM} $ (i.e. a transaction between two members who never made a transaction before) and the closest matching certification in $ \mathcal{C} $. \textbf{Bottom} - Distribution of the number of matching transactions in $ \mathcal{T}_{MM} $ preceding a certification in $ \mathcal{C} $ for certifications occurring after prior transactions.}
\label{fig:time_to_wait}
\end{center}
\end{figure}

When a certification occurs between two previously disconnected nodes in $\mathcal{C}$ we search for the first transaction between these two nodes. This transaction can happen before the certification, after, or never. We observe that for $73\%$ of certifications, the two involved nodes never make a transaction between them. $16 \% $ of certifications have a pre-existing transaction between the nodes, and $11 \%$ will make a transaction in the future. The top plot of Figure \ref{fig:time_to_wait} shows the distribution of the time seperating a certification between two nodes in $ \mathcal{C} $ from the closest matching transaction between these two nodes in $ \mathcal{T}_{MM} $. Note that most new certification relationships which have a matching transaction in $\mathcal{T}_{MM} $ are relatively close in time. One possible explanation for this is that new members are often involved in small transactions (welcome gifts or acknowledgments) shortly after (or before) being certified. The bottom plot of figure \ref{fig:time_to_wait} shows the distribution of the number of transactions in $ \mathcal{T}_{MM} $ preceding a certification in $ \mathcal{C} $ for these certifications that occur after some number of prior transactions. Almost all such certifications occur after only one or two transactions, but in a very few cases, a certification can happen after as many as $14$ prior transactions.

% OBSERVATION 5
%
\begin{observation}
Most certification relationships in $ \mathcal{C} $ do not have a matching transaction in $\mathcal{T}_{MM} $, and when they do, they tend to be close in time.
\label{obs:5}
\end{observation}

% FIGURE 7
%

Now, we look at this the other way around. When a transaction occurs in $ \mathcal{T}_{MM} $, are the two involved nodes already linked by a certification or will they certify in the future? $64\%$ of all transactions from $ \mathcal{T}_{MM} $ occur between two members linked by a certification relationship in $\mathcal{C}$. More precisely, $42\%$ occur between two already certified members, while only $22\%$ occur between members who will certify themselves in the future. These numbers also mean that about $36\%$ of all transactions from $ \mathcal{T}_{MM} $ occurs between members who never certify themselves directly in $\mathcal{C}$. We can also consider only new transactions in $ \mathcal{T}_{MM} $, that is transactions between two members who never made a transaction before. The middle plot of figure \ref{fig:time_to_wait} shows the distribution of time intervals $ \Delta t $ (in seconds) between a new transaction in $ \mathcal{T}_{MM} $ and the closest matching certification in $ \mathcal{C} $. As can be seen, the majority of new transactions in $ \mathcal{T}_{MM} $ occurs shortly after a matching certification in $ \mathcal{C} $.

% OBSERVATION 6
%
\begin{observation}
Most transactions in $\mathcal{T}_{MM} $ occur between members linked by a certification in $\mathcal{C}$.
\label{obs:6}
\end{observation}

Because of the constraints imposed on the WoT, the undirected version of $ G \left( \mathcal{C} \right) $ is connected and we can compute a certification distance between any two members. Because the WoT results from social interactions, we expect that nodes lying far appart in $ G \left( \mathcal{C} \right) $ have a small probability of knowing each other, and therefore, of making transactions. About $ 78\% $ of the transactions between non-certified members occur between members at distance $2$ in $ G \left( \mathcal{C} \right) $, $ 19\% $ at distance $3$, $ 2.8\% $ at distance $4$, and $ 0.2\% $ at distance $5$, such that the more distant two nodes are in $ G \left( \mathcal{C} \right) $, the less likely they are to make transactions in $ \mathcal{T}_{MM} $.

% OBSERVATION 7
%
\begin{observation}
Members who make transactions in $\mathcal{T}_{MM}$ without being certified in $\mathcal{C}$ are more likely to be close in $ G \left( \mathcal{C} \right) $.
\label{obs:7}
\end{observation}

The probability of having a link between two randomly chosen nodes in $ \mathcal{C}$ and $ \mathcal{T}_{MM} $ are, by definition, the densities of the induced graphs $ G \left( \mathcal{C} \right) $ and $ G \left( \mathcal{T}_{MM} \right) $. Two randomly chosen nodes have a certification link with probability $6.15 \times 10^{-3}$ while they made at least one transaction with probability $ 4.67 \times 10^{-3} $. Now, if instead of picking these nodes randomly, we sample nodes sharing a certification link, the probability of making a transaction increases to $0.24$. In the same way, sampling among nodes having at least one transaction connecting them increases the proability of certification to $0.53$.

% OBSERVATION 8
%
\begin{observation}
The probability two nodes to be linked in one stream is orders of magnitudes larger if they are linked in the other stream.
\label{obs:8}
\end{observation}

The observations from this section lead to think that certifications are often the first kind of link appearing between two unrelated nodes in $\tilde{G}1$, and that transactions tend to occur afterwards. Moreover, we showed that these transactions do not happen randomly but constrained by the social network. This means that transactions preferably take place between friends or between friends of friends than between strangers (i.e. nodes far away in the WoT). In the next section, we will investigate these questions further using more complex concepts mixing time and structure.

%%%%%%%%%%%%%%%%%%%%%%%%%%%%%%%%%%%%%%%%%%%%%%
%      SECTION VI - NEIGHBORHOODS IN C AND T
%%%%%%%%%%%%%%%%%%%%%%%%%%%%%%%%%%%%%%%%%%%%%%
%
\section{Member-centric analysis of certifications and transactions}
\label{sec:neighborhoods} 

In the previous section, we studied how new links where occurring in $ \mathcal{C} $ and $ \mathcal{T}_{MM} $ using a simple link-based approach. In this section, we wish to go further and exploit the topology of both streams to gain more insights. A widely used topological concept in graph theory is the one of neighborhoods and egonets. For a stream graph $ S = (T,V,W,E) $ the definition of the neighborhood of node $v \in V$, as given in \cite{stream_graphs}, is the following cluster:

\begin{equation}
N_S \left(v \right) = \left\{ \left( t,u \right), \left( t, uv \right) \in E \right\}
\label{eq:neighborhood}
\end{equation}

A neighborhood is thus composed of temporal nodes rather than simple nodes for static graphs. We also consider the neighborhood of the induced graph $ G(S) $:

\begin{equation}
\bar{N}_{S} \left(v \right) = \left\{ u \in V, \exists \left( t, uv \right) \in E \right\}
\label{eq:aggregated_neighborhood}
\end{equation}

In simple words, $ \bar{N}_{S} \left(v \right) $ contains all nodes which interacted with $v \in V$ at least once, while $ N_S \left( v \right) $ keeps track of the interaction times.

%%%%%%%%%%%%%%%%%%%%%%%%%%%%%%%%%%%%%%%%%%%%%%%%%%%%%%%%%%%%%%%
%      VI.1 - DO MEMBERS TRADE WITH CERTIFICATION NEIGHBORS ?
%%%%%%%%%%%%%%%%%%%%%%%%%%%%%%%%%%%%%%%%%%%%%%%%%%%%%%%%%%%%%%%
%
\subsection{Do members trade with their certification neighbors or vice-versa?}
\label{subsec:neighbors}

The first question which comes to mind is whether nodes tend to have similar neighborhoods in $\mathcal{C} $ and $ \mathcal{T}_{MM} $. In other words whether users tend to be friends with their transaction partners or make transactions with their friends. In order to compare the neighborhoods, we use the Jaccard index and the overlap coefficient defined between two sets $a$ and $b$ as :

\begin{equation}
J\left( a,b \right) = \frac{\left| a \cap b \right|}{\left| a \cup b \right|}, \ \ \ \ \ \ \ \ \ \ \ O \left( a,b \right) = \frac{\left| a \cap b \right|}{min \left( \left| a \right|, \left| b \right| \right)}
\label{eq:jaccard_overlap}
\end{equation}

% FIGURE 8
%
\begin{figure}
\centering
\includegraphics[scale=.32]{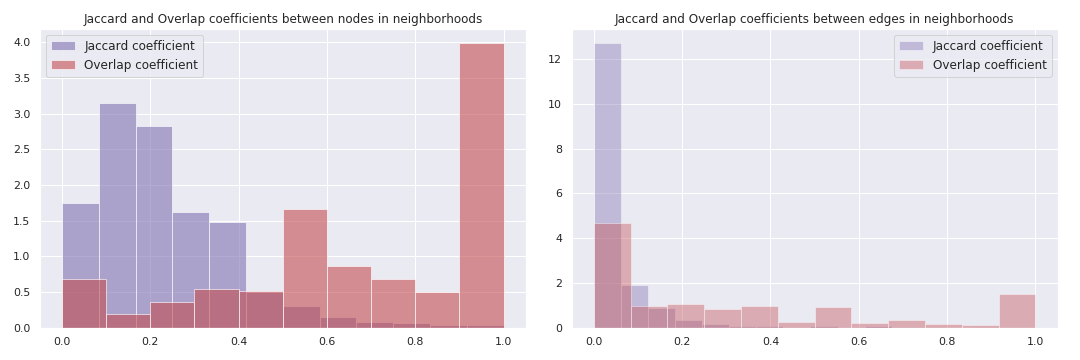}
\caption{Distributions of the Jaccard coefficient (in blue) and the Overlap coefficient (in red) between \textbf{Left -} the nodes' transaction neighborhoods $\bar{N}_{\mathcal{T}_{MM}} \left( u \right) $ and certification neighborhoods $\bar{N}_{\mathcal{C}} \left( u \right) $ \textbf{Right -} the edges within these neighborhoods.}
\label{fig:jaccard-overlap}
\end{figure}

The left plot of figure \ref{fig:jaccard-overlap} shows the distribution of the Jaccard coefficient over the nodes (in blue) and the distribution of the overlap coefficient (in red) between $ \bar{N}_{\mathcal{T}_{MM}} \left( u \right) $ and $ \bar{N}_{\mathcal{C}} \left( u \right) $. As can be seen a large majority of nodes have a strong overlap between their two neighborhoods, but rather small values for the jaccard coefficients. This can be explained by the fact that the transaction neighorhood is smaller than the certification neighborhood for about $90\%$ of the nodes and tends to be mostly included within the certification neighborhood. This means that members are more inclined to make transactions with members within their certification neighborhood, but tend to certify people they never make transaction with, which is in accordance with the link-based results of the previous section.

% OBSERVATION 9
%
\begin{observation}
Members tend to make transactions mostly with people within their certification neighborhood, but tend to certify people they never make transactions with.
\label{obs:9}
\end{observation}

The right plot of figure \ref{fig:jaccard-overlap} shows the distributions of the same two coefficients between the edges within $ \bar{N}_{\mathcal{T}_{MM}} \left( u \right) $ and $ \bar{N}_{\mathcal{C}} \left( u \right) $. As can be seen, even though nodes tend to make transactions with people they are linked to in $ \mathcal{C} $ there is a much smaller overlap between the interactions in these neighborhoods.

%%%%%%%%%%%%%%%%%%%%%%%%%%%%%%%%%%%%%%%%%%%%%%%%%%%%%
%      VI.2 - CLUSTERING, TRIANGLES, AND K-CLOSURES
%%%%%%%%%%%%%%%%%%%%%%%%%%%%%%%%%%%%%%%%%%%%%%%%%%%%%
%
\subsection{Clustering, triangles, and k-closures}
\label{subsec:clustering}

In addition to neighborhoods and ego-nets, we propose to study triangles in $ \mathcal{C}$ and $\mathcal{T}_{MM} $, which are a famous and very important concept in network theory. They intuitively convey the idea that \textit{people I interact with tend to also interact between themselves}. Two basic related metrics in graphs are the clustering coefficient and the number of triangles. 

The average clustering coefficent of $G \left( \mathcal{C} \right) $ (resp. $G \left( \mathcal{T}_{MM} \right)$) is $0.49$ (resp. $0.31$), while the average clustering of $ G \left( \mathcal{T}_{AA} \right) $ is $0.13$. Note that, by construction, the clustering coefficients for $ G \left( \mathcal{T}_{AM} \right) $ and $ G \left( \mathcal{T}_{MA} \right) $ are both equal to 0. In terms of triangles, $G \left( \mathcal{C} \right) $ contains $6589$ triangles, $G \left( \mathcal{T}_{MM} \right)$ $1990$ triangles, and $ G \left( \mathcal{T}_{AA} \right) $ only $393$.

% FIGURE 9
%
\begin{figure}
\begin{center}
\includegraphics[scale=.6]{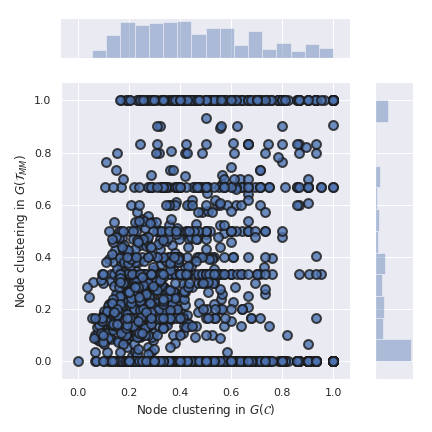}
\end{center}
\caption{Node clustering distributions in $ G \left( \mathcal{C} \right)$ and $ G \left( \mathcal{T}_{MM} \right) $.}
\label{fig:clustering}
\end{figure}

First of all, note that these values are much higher than for random graphs with the same degree distributions (about 23 times higher than the average for $G \left( \mathcal{C} \right) $, and about 4 times higher than average for $ G \left( \mathcal{T}_{MM} \right) $), meaning that both $G \left( \mathcal{C} \right) $ and $ G \left( \mathcal{T}_{MM} \right) $ are very clustered networks. Figure \ref{fig:clustering} shows the distributions of the nodes' clustering coefficient values in both networks. Note that, although their average clustering values are similar, the distributions over the nodes are very different. Indeed, the distribution for $ G \left( \mathcal{T}_{MM} \right) $ exhibits two peaks around 0 and 1, meaning that most nodes make transactions either in a very clustered way or not at all, while the distribution for $G \left( \mathcal{C} \right) $ has a majority of values in between those extremes.

% OBSERVATION 10
%
\begin{observation}
Both $ G \left( \mathcal{C} \right) $ and $ G \left( \mathcal{T}_{MM} \right) $ display a strongly clustered structure. However, the node clustering distributions have very different shapes in each stream. The node clustering distribution of $ G \left( \mathcal{T}_{MM} \right) $ exhibits two peaks around 0 and 1, meaning that members tend to make transactions either in a very clustered way or not at all.
\label{obs:10}
\end{observation}

The previous results are built on the induced graphs of both streams such that we do not take time into account. The clustering coefficient has been extended to stream graphs (see \cite{stream_graphs}), both in continuous and discrete time, but will provide little insight here. Indeed, the slow dynamics of $ \mathcal{C} $ makes the results very similar to the ones obtained for the induced graph $G \left( \mathcal{C} \right) $, and $\mathcal{T}_{MM} $ is a very sparse stream such that there is almost no time instant where transactions form triangles. Note that a rather artificial way to compute stream clustering values for $\mathcal{T}_{MM} $ would be to set a sufficiently large link duration in continuous time such that transactions from different blocks of the blockchain start to overlap.

\begin{figure}
\centering
\includegraphics[scale=.32]{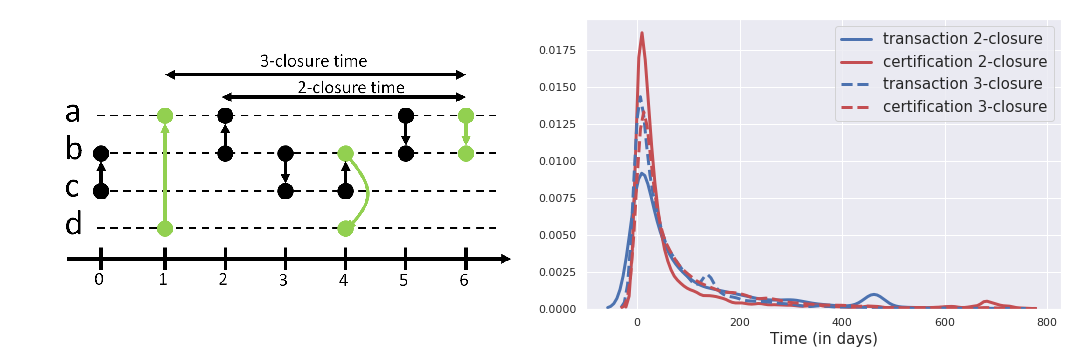}
\caption{\textbf{Left -} Schematic view of the 2 and 3-closures. The 2-closure of link $(6,ab)$ is equal to $4$ since we have to go back to $t=2$ to find a reversed link (e.g. $(2,ba)$). The 3-closure of link $(6,ab)$ is equal to $5$ since we have to go back to $t=1$ to close the triangle composed of $(6,ab), (4,bd), (1,da)$. \textbf{Right -} The densities of 2 and 3 closures for all links in $\mathcal{C}$ and $ \mathcal{T}_{MM} $. Note that these only consider links with finite k-closures, links with infinite values are ignored.}
\label{fig:closures}
\end{figure}

Here, we prefer to use another stream graph concept to study triangles with explicitely taking time into account: the k-closures, where k is an integer larger than 1. The 2-closure, for example, is defined for a given link, as the amount of time we have to go back to find a link beween the same nodes but in the opposite direction. The 3-closure is defined as the amount of time required to find a triangle containing the link considered. The left plot of figure \ref{fig:closures} shows a schematic view of the 2 and 3-closures in a simple link stream. The 2-closure of link $(6,ab)$ for example is equal to $4$ since we have to go back to $t=2$ to find a link in the opposite direction, namely link $(2,ba)$. The 3-closure of the same link $(6,ab)$ is equal to $5$ since, this time, we have to go back to $t=1$ to find the triangle $ (6,ab),(4,bd),(1,da)$. Note that the k-closure values can be infinite if there is no link in the opposite direction in the past, or no way to close the triangle.

Here we examine both the 2 and 3-closures for $ \mathcal{C} $ and $ \mathcal{T}_{MM}$. The right plot of figure \ref{fig:closures} shows the k-closure densities for the two streams. As can be seen, the 2-closures for $\mathcal{C}$ tend to be smaller than for $ \mathcal{T}_{MM} $, meaning that people tend to certify back faster than they make a transaction in the other direction. The 3-closures densities of $ \mathcal{C} $ and $ \mathcal{T}_{MM}$ are very similar which seems to suggest that, although there are less triangles in $ \mathcal{T}_{MM}$ than in  $ \mathcal{C} $, they tend to appear within the same time frame.

% OBSERVATION 11
%
\begin{observation}
Users tend to certify back faster than they make a backward transaction. On the other hand, triangles are built within similar time frames in $\mathcal{C}$ and $\mathcal{T}_{MM} $.
\label{obs:11}
\end{observation}

% FIGURE 10
%

%%%%%%%%%%%%%%%%%%%%%%%%%%%%%%%%%%
%      CONCLUSION
%%%%%%%%%%%%%%%%%%%%%%%%%%%%%%%%%%
%
\section{Conclusion}
\label{sec:conclusion}

We studied a recent cryptocurrency which, despite its relatively small size, provides a very interesting and publicly available dataset to study transactions within a social environment. Because the data has both a structural and temporal component, neither a pure static graph nor a time serie based model are well suited for its study. We proposed here to rely on the stream graph formalism to study the streams of certifications and transactions. 

We showed that certifications are often the first type of link occurring between two previously disconnected nodes, suggesting that members of $\tilde{G}1$ start to meet at social events before making transactions. In this context, the social network shapes the way transactions occur: even when they occur between two members without a certification relationship, this transaction has a much higher probability of occurring between two socially close nodes. Investigating further, we showed that members tend to have a transaction neighborhood included within their certification neighborhood, meaning that users make transactions mostly with their acquaintances, but with a subset of them only.

Finally, we studied the clustering of both the certification and transaction streams and discovered that, although both streams display a rather high average clustering coefficient, their node clustering distributions have very different shapes. Looking at the distribution for the transaction stream, it seems like nodes tend to make transactions either in a very clustered way or not at all, but rarely in between these two extremes. We also studied the k-closures of certifications and transactions and found that users tend to certify back faster than they tend to make a backward transaction, meaning that certification relationships seem to become bi-directional faster than transactions.

We want to stress out that these results might be specific to this particular cryptocurrency and should not be considered as general principles. Indeed, $\tilde{G}1$ is still very young and used mostly by currency-interested people, such that most transactions occur, for the moment, during special social events where members meet and welcome new users. As the currency will age, we expect that transactions will occur in a less socially confined way (through exchange platforms for example), and that time will play a more central role in the study of the interplay between social ties and interactions. 

Several directions for future work exist. First, more advanced stream graph concepts may help understanding the structure of blockchain based systems better. Another interesting direction would be to compare the properties of $\tilde{G}1$ with other social systems with transactions, like for instance Steemit \cite{Steem}. We believe that these blockchain based systems will become more and more present and will provide fascinating data sources for such studies.

\subsection*{Acknowledgements}

The authors would like to thank Hugo Trentesaux for his help.
This work is funded in part by the ANR (French National Agency of Research) under the FiT LabCom grant.

\bibliography{article} 
\bibliographystyle{plain}

\end{document}